\documentclass[aps,prl,reprint,superscriptaddress]{revtex4-2}

\usepackage[english]{babel}
\usepackage{graphics}
\usepackage{array}
\usepackage{graphicx}
\usepackage{todonotes}
\usepackage{amsfonts,amssymb,amsmath}
\usepackage{color}
\usepackage{threeparttable}
\usepackage{physics}

\usepackage[separate-uncertainty,per-mode=fraction]{siunitx}

\definecolor{darkblue}{rgb}{0.0,0.0,0.4}
\definecolor{darkgreen}{rgb}{0.0,0.4,0.0}
\definecolor{darkred}{rgb}{0.6,0.0,0.0}
\usepackage[bookmarksopen]{hyperref}
\hypersetup{colorlinks,linkcolor=darkblue,citecolor=darkblue,urlcolor=darkblue}

\newcommand{\baf}{${}^{138}\mathrm{BaF}$}
\newcommand{\gs}{$\mathrm{X}^2\Sigma^+$}
\newcommand{\exs}{$\mathrm{A}^2\Pi_{1/2}$}
\newcommand{\baff}[1]{${}^{#1}\mathrm{BaF}$}

\newcolumntype{M}[1]{>{\centering\arraybackslash}m{#1}}

\begin{document}

\title{Laser cooling of barium monofluoride molecules\\ using synthesized optical spectra}

\author{Marian Rockenh\"auser}
\thanks{These two authors contributed equally}
\affiliation{5. Physikalisches  Institut  and  Center  for  Integrated  Quantum  Science  and  Technology, Universit\"at  Stuttgart,  Pfaffenwaldring  57,  70569  Stuttgart,  Germany}

\author{Felix Kogel}
\thanks{These two authors contributed equally}
\affiliation{5. Physikalisches  Institut  and  Center  for  Integrated  Quantum  Science  and  Technology, Universit\"at  Stuttgart,  Pfaffenwaldring  57,  70569  Stuttgart,  Germany}

\author{Tatsam Garg}
\affiliation{5. Physikalisches  Institut  and  Center  for  Integrated  Quantum  Science  and  Technology, Universit\"at  Stuttgart,  Pfaffenwaldring  57,  70569  Stuttgart,  Germany}

\author{Sebasti\'an A. Morales-Ram\'irez}
\affiliation{5. Physikalisches  Institut  and  Center  for  Integrated  Quantum  Science  and  Technology, Universit\"at  Stuttgart,  Pfaffenwaldring  57,  70569  Stuttgart,  Germany}

\author{Tim Langen}
\email{t.langen@physik.uni-stuttgart.de}

\affiliation{5. Physikalisches  Institut  and  Center  for  Integrated  Quantum  Science  and  Technology, Universit\"at  Stuttgart,  Pfaffenwaldring  57,  70569  Stuttgart,  Germany}

\affiliation{Vienna Center for Quantum Science and Technology, Atominstitut, TU Wien,  Stadionallee 2,  A-1020 Vienna,  Austria}

\begin{abstract}
We demonstrate laser cooling of barium monofluoride (\baff{138}) molecules. We use serrodynes to synthesize time-sequenced optical spectra that can be precisely tailored to the hyperfine structure of this heaviest non-radioactive alkaline earth monofluoride. By optimizing these optical spectra, we realize strong Sisyphus cooling forces that efficiently collimate a molecular beam. Our technique is an important step towards using intense beams of barium monofluoride for precision measurement applications, and will be useful for cooling other molecular species with complex level structure.
\end{abstract}

\maketitle


Laser cooling of molecules has made remarkable progress over the last years~\cite{Fitch2021,Langen2023}, and a wide variety of molecular species from diatomics~\cite{Barry2011,Barry2014,Truppe2017,Anderegg2017,Collopy2018,Lim2018,McNally2020,VazquezCarson2022} to  polyatomics~\cite{Vilas2021,Kozyryev2017,Augenbraun2020,Mitra2020} can now be routinely cooled. 

Recently, significant efforts have been made to add barium monofluoride (BaF)  to the list of laser-coolable species. 
This species is promising for searches of the electron's permanent electric dipole moment
~\cite{Aggarwal2018,Vutha2018}, precision tests of weak parity violation~\cite{Demille2008,Altuntas2018}, studies of molecular Rydberg and plasma physics~\cite{Zhou2015,Marroquin2024}, as well as the simultaneous cooling of isotopic mixtures of molecules~\cite{Tomza2015,Kogel2021}. Due to the high mass of BaF, laser cooling of this species would also be an important step towards cooling even heavier, short-lived species, such as radium monofluoride~\cite{Isaev2010,GarciaRuiz2020}, which show exceptional sensitivity to symmetry-violating nuclear properties~\cite{Arrowsmithkron2023}.

 Previous experimental work on BaF focused on buffer gas cooling~\cite{Bu2017,Albrecht2020}, ro-vibrational cooling~\cite{Cournol2018,Courageux2022}, electrostatic focusing~\cite{Touwen2024}, optical cycling~\cite{Albrecht2020}, deflection~\cite{Chen2017} and trapping in a matrix~\cite{Li2023}. However, as a consequence of the high mass, a comparably narrow linewidth~\cite{Aggarwal2019} and resolved excited state hyperfine structure~\cite{Denis2022,Marshall2022,Bu2022,Rockenhaeuser2023}, as well as potential branching losses through an intermediate electronic state~\cite{Hao2019}, laser cooling of BaF has so far remained elusive. While evidence for Doppler forces has recently been reported~\cite{Zhang2022}, in the corresponding experiments branching into unaddressed states resulted in most molecules being lost, rather than the gas being cooled. 

Here, we realize efficient Sisyphus laser cooling of a \baf\, molecular beam. To address the resolved hyperfine structure of BaF, we drive the molecular cycling transitions using optimized optical spectra, which we synthesize using serrodynes~\cite{Rogers2011,Holland2021}. This technique is more versatile and power-efficient than the conventional laser sidebands that are commonly used in molecular laser cooling, and enables the realization of strong laser cooling forces using only moderate experimental resources. As an added benefit, the approach greatly simplifies the experimental setup by eliminating the need for free-space optics to prepare the cooling laser frequencies. 

\begin{figure}[tb]
\centering
\vspace{4pt}
\includegraphics[width=0.46\textwidth]{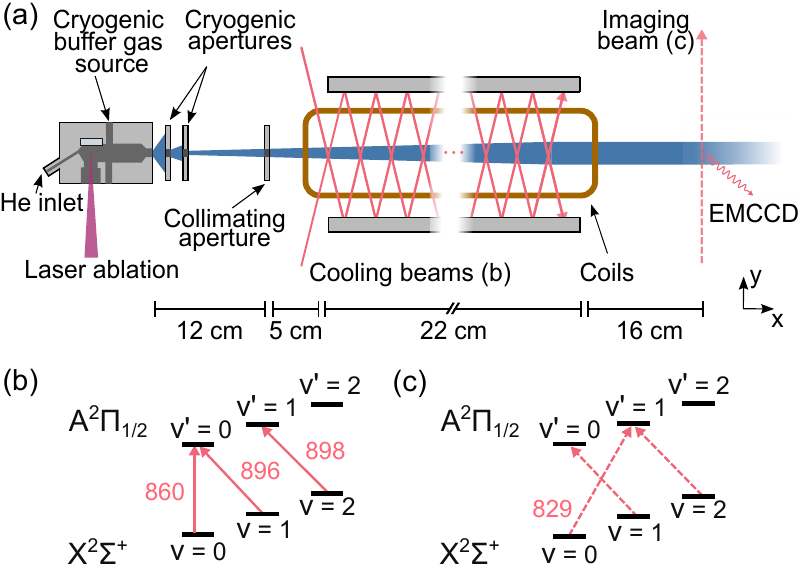}
 \caption{(a) Experimental setup. BaF molecules are created using laser ablation in a cryogenic buffer gas source. After leaving the source, they form a molecular beam that is collimated by three apertures, the last of which is located $12\,$cm downstream from the buffer gas cell. The beam then enters a $22\,$cm long laser cooling region, where two cooling laser beams with opposite propagation directions are superimposed and retro-reflected with $17$ round-trip passes to form a standing intensity wave pattern. Coils generate a homogeneous magnetic field in this region that points into the image plane and remixes molecular dark states. After a short further time-of-flight, molecular fluorescence is imaged using an EMCCD camera. Dimensions are not to scale. (b,c) Vibrational level scheme for cooling (b) and Raman optical cycling imaging (c)~\cite{Rockenhaeuser2023}. Arrows indicate the transitions addressed by cooling (solid) and imaging beams (dashed), respectively, with numbers denoting the corresponding wavelengths in nanometers.}
\label{fig:experimental_setup}
\end{figure}

A summary of the experimental setup is depicted in Fig.~\ref{fig:experimental_setup}. 
The experiment starts with the creation of BaF molecules by laser ablation inside a cryogenic buffer gas cell. The hot ablated molecules thermalize with a helium buffer gas and form a divergent molecular beam with a forward velocity of around $170\,$m/s~\cite{Albrecht2020}. This beam is collimated transversally using three apertures, the last of which limits the transverse velocity spread to around $\pm2.7\,$m/s full width at half maximum, or, equivalently, a transverse temperature of around $100\,$mK. The molecules subsequently enter a $22\,$cm long laser cooling region, where they interact for approximately $1.3$\,ms with a pair of retro-reflected laser beams with parallel linear polarizations. These beams form an intensity standing wave, and contain suitable cooling and vibrational repumping light with a peak intensity well above saturation (see Fig.~\ref{fig:experimental_setup}b and Appendix). Throughout the cooling region, a set of coils allows
for the application of a magnetic field to remix molecular dark states~\cite{Albrecht2020}.

Following cooling and subsequent free flight, laser-induced fluorescence (LIF) of the molecules is imaged onto an EMCCD camera using Raman optical cycling imaging~\cite{Rockenhaeuser2023,Shaw2021}. This technique inherently includes the relevant vibrational repumping lasers (Fig.~\ref{fig:experimental_setup}c) and therefore, in contrast to regular LIF detection employed in other molecular laser cooling experiments, does not require a vibrational cleanup region prior to imaging. 

\begin{figure}[tb]
\centering
\includegraphics[width=0.46\textwidth]{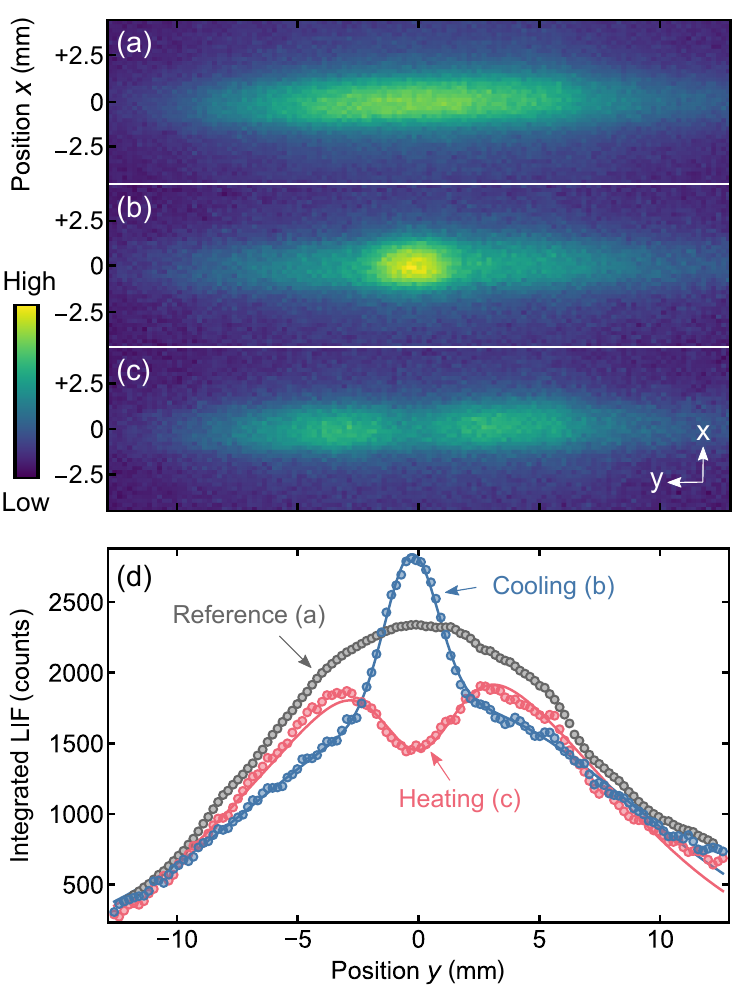}
 \caption{Laser-induced fluorescence (LIF) images of \baff{138} molecules. The images are averaged over $120$ experimental realizations and color denotes LIF signal strength. (a) Unperturbed reference with no laser applied in the cooling region. (b) Sisyphus cooling for blue-detuned cooling laser. (c) Sisyphus heating for red-detuned cooling laser. (d) Integrated line profiles of the data in (a,b,c). Solid lines are fits using a double-Gaussian function (see Appendix). Cooling is realized using a three-frequency serrodyne waveform (scheme I in Fig.~\ref{fig:levelscheme}) and optimized parameters discussed in the Fig.~\ref{fig:simulations}.}
\label{fig:cooling}
\end{figure}

Example images of the molecular beam, highlighting the effects of both Sisyphus cooling for blue-detuned, and heating for red-detuned cooling laser light are shown in Fig.~\ref{fig:cooling}. In the cooling configuration, we observe a significantly enhanced population in the coldest part of the molecular distribution, yielding a well-collimated molecular beam. As expected from the low Sisyphus cooling limit of about $10\,\mu$K, very little transverse thermal expansion remains and the size of this coldest part of the beam is comparable to the final collimating aperture. Assuming initially homogeneous spatial and Gaussian velocity distributions at the buffer gas cell exit, a conservative upper limit for the temperature can thus be estimated from the data to be well below $2.5\,$mK~\cite{Kozyryev2017,McNally2020,Augenbraun2020,VazquezCarson2022}. Taken together, this highlights the efficiency of the laser cooling and the low loss rates achievable in BaF,  which are competitive with the best results previously obtained for other species~\cite{Shuman2010,Lim2018,Kozyryev2017,Augenbraun2020,Mitra2020}. 

We now investigate the origin, implementation and properties of the cooling forces in more detail. Optical cycling and laser cooling of BaF rely on the diagonal Franck-Condon factors of this species~\cite{Chen2016,Albrecht2020,Rockenhaeuser2023}. For cooling we employ the \gs ($\nu=0$) $\rightarrow$ \exs ($\nu'=0$) transition, where $\nu$ is the vibrational quantum number. In order to address the residual vibrational branching, we use two repumping lasers resonant with the \gs ($\nu=1$) $\rightarrow$ \exs ($\nu'=0$) and \gs ($\nu=2$) $\rightarrow$ \exs ($\nu'=1$) transitions, respectively (see also Fig.~\ref{fig:experimental_setup}b). Taken together, this allows the scattering of thousands of photons, before molecules will be lost into higher vibrational levels with $\nu\geq3$ or intermediate electronic states~\cite{Hao2019}.

\begin{figure}[tb]
\centering
\includegraphics[width=0.455\textwidth]{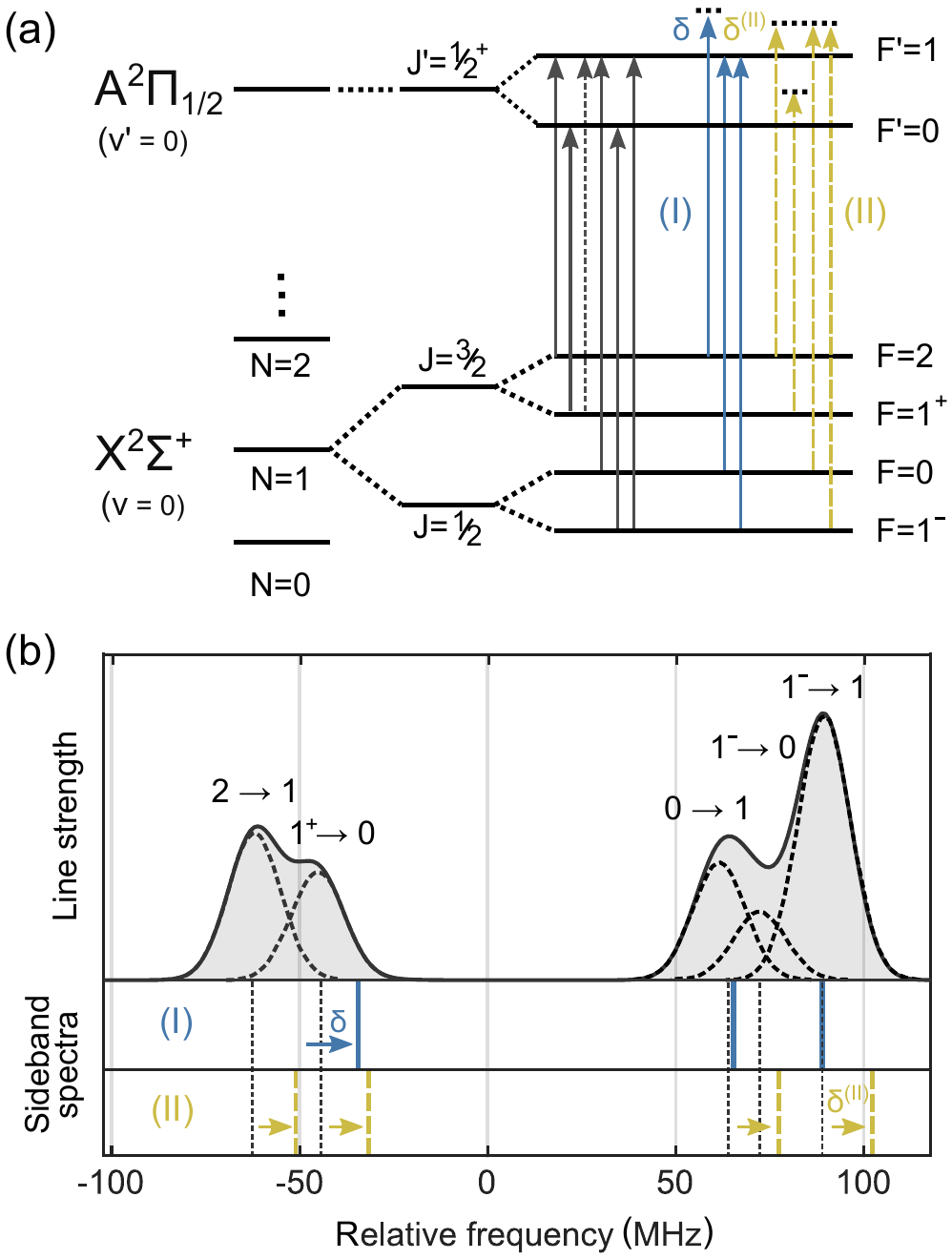}
 \caption{(a) Rotational and hyperfine structure in \baff{138}. The hyperfine levels of the excited \exs\, state are resolved, resulting in six allowed transitions (arrows), with one (dotted arrow) being negligibly weak. Addressing of these transitions can be accomplished using three optimized sidebands (scheme I, blue solid arrows), or using four more conventional sidebands (scheme II, yellow dashed arrows). In the former, the sideband addressing the $F=2 \rightarrow F'=1$ and $F=1^+ \rightarrow F'=0$  transitions is detuned by a variable detuning $
 \delta$, while in the latter all four sidebands are simultaneously detuned by $
 \delta^{(II)}$. (b) Comparison of the two schemes to a simulated reference spectrum.}
\label{fig:levelscheme}
\end{figure}

In order to close the optical cycle rotationally, we follow the usual strategy~\cite{Stuhl2008} of tuning cooling and repumping lasers such that they address transitions from the respective $N=1$ ground states to $J^p=1/2^+$ excited states. Here, $N$ denotes the rotational quantum number in the ground state, $J$ the total angular momentum in the excited state and $p=+$ in the exponent a state with positive parity. The corresponding level structure and transitions are summarized in Fig.~\ref{fig:levelscheme}a. 

In BaF these rotational states split further due to spin-rotation and hyperfine interactions. For the \gs ground state this leads to four almost equidistant hyperfine levels with $F=1^-,0,1^+,2$, where $F$ is the hyperfine quantum number. Here, we employ the common notation with $+$ and $-$ to distinguish the two levels with $F=1$~\cite{Rockenhaeuser2023}, which should not be confused with the parity of the excited state. Typically, sidebands must be applied to the cooling laser in order to address all of these ground state hyperfine levels.

For the \exs\, excited state, it was initially assumed that the $F'=0,1$ hyperfine levels were degenerate~\cite{Chen2016,Albrecht2020}. However, recent work has shown that the splitting between the $F'=0$ and $F'=1$ states is about $17\,$MHz~\cite{Denis2022,Marshall2022,Bu2022,Rockenhaeuser2023}. Given the excited state linewidth of $2.84\,$MHz, the splitting is thus, in contrast to the more common alkaline earth monofluorides CaF and SrF, well resolved and needs to be taken into account when applying sidebands to the cooling laser~\cite{Rockenhaeuser2023}. 

\begin{figure}[tb]
\includegraphics[width=0.45\textwidth]{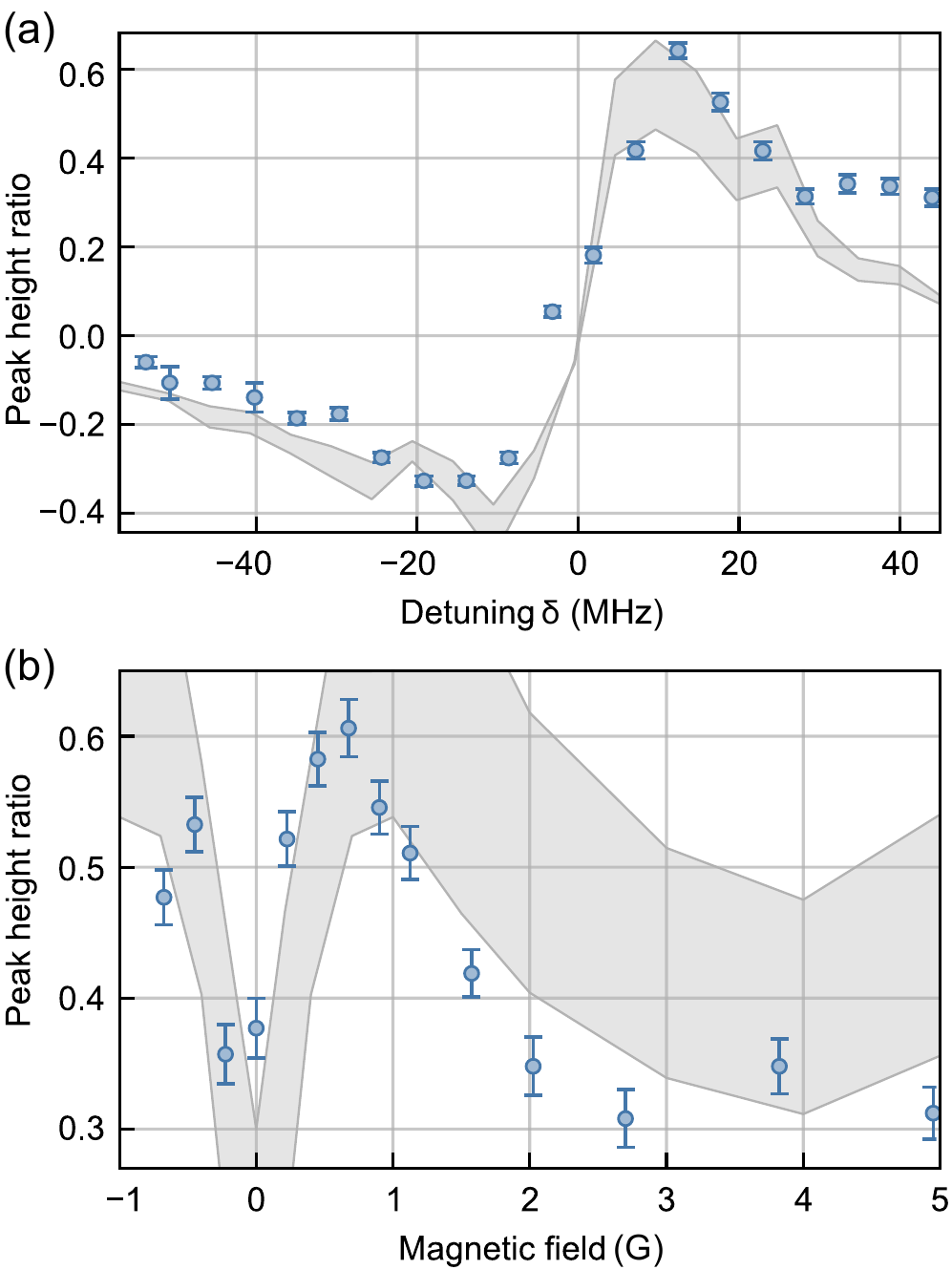}
 \caption{Characterization of the laser cooling process as a function of (a) detuning and (b) magnetic field. Initial powers in the cooling region are $60\,$mW per beam for the cooling light and $45\,(15)\,$mW for the first (second) vibrational repumper for all measurements. (a) highlights how Sisyphus laser cooling forces are active over a wide range of detunings, with maximum cooling occurring at $\delta=12.5\,$MHz, and maximum heating at $-16\,$MHz. Data were taken with a magnetic field of $0.7\,$G. In (b) we use a detuning of $\delta=12\,$MHz. Individual data points were averaged over $30$ experimental realizations. Theory predictions (shaded areas) are based on Monte Carlo simulations and solutions of optical Bloch equations.}
\label{fig:simulations}
\end{figure}
In order to identify the most favorable sideband configuration, we perform simulations based on multi-level optical Bloch equations~\cite{Devlin2018,Kogel2021,Kogel2021code}. In general, laser cooling forces in molecules are a combination of Doppler- and Sisyphus-type forces. For our parameters, however, recoil and scattering rates are small and we find that Doppler cooling and heating are negligible, especially for the transverse velocity range present in our molecular beam. Instead, the cooling is dominated by magnetically-assisted Sisyphus forces~\cite{Emile1993}. 

While the usual procedure in molecular laser cooling is to address all ground state hyperfine levels using an identical number of sidebands, we find it to be beneficial to apply only three sidebands to address the four ground state hyperfine levels in BaF (scheme I in Fig.~\ref{fig:levelscheme}). In the most favorable configuration the $F=2 \rightarrow F'=1$ and $F=1^+ \rightarrow F'=0$ transitions are addressed with the strongest sideband, which is blue-detuned by approximately $12.5\,$MHz, or $4.4$ linewidths. Here, $\delta=0$ is chosen such that it corresponds to the frequency between both transitions, where Sisyphus cooling and heating cancel each other. This is combined with a near-resonant addressing of the $F=0\rightarrow F'=1$ and $F=1^-\rightarrow F'=1$ transitions, respectively. The result is reminiscent of $\Lambda$-enhanced molasses schemes previously used to cool CaF molecules~\cite{Caldwell2019,Li2023}, which share a very similar level structure. It is also in line with our earlier, more general observation that it can be beneficial if only a subset of the transitions of a molecule are addressed with high powers to generate the laser cooling forces, while the rest are only driven off-resonantly to avoid dark states~\cite{Kogel2021}. 


In order to synthesize the optical spectra for the cooling transition in the most flexible manner, we use serrodynes to generate them ~\cite{Rogers2011,Holland2021}. This technique uses linear phase shifts applied using an electro-optic modulator (EOM) to generate frequency components that are shifted from the carrier frequency of a laser beam. It has previously been used to shift frequencies both in the microwave~\cite{Cumming1957} and optical domain~\cite{Johnson2010}.

In our experiment, the linear phase shifts for the cooling light are created by applying a linear voltage ramp to a high-bandwidth fiber-coupled EOM, using the amplified output of an arbitrary waveform generator with a bandwidth of up to around $250\,$MHz. The use of serrodynes --- where ideally only one frequency component is present at a given time --- allows us to conveniently place the fiber-coupled EOM between the diode laser and tapered amplifier used in our setup, thus eliminating any free space optics in the preparation of the cooling light. 

By concatenating linear voltage ramps with different slopes, well-defined time sequences of frequency components with specific shifts can be generated~\cite{Rogers2011}. These components can be tuned independently, resulting in nearly arbitrary optical spectra. These spectra can be tailored to the individual hyperfine transitions between the ground and excited states in the most favorable way. 

Notably, for our parameters we find that only moderate switching times between the individual frequency components --- well above the timescale set by the linewidth of the transition --- are required to saturate the laser cooling forces. We attribute this to the presence of residual higher-order frequency components which permanently repump all hyperfine states (see Appendix), as well as the fact that Sisyphus forces strongly increase with power, rather than scattering rate. The concentration of nearly all available power in a single frequency sideband at any given time thus over-compensates for any potential loss in scattering rate that may result from too slow switching. A detailed comparison between serrodynes and conventional sidebands will be presented in future work. 

In Fig.~\ref{fig:simulations} we experimentally characterize the transverse laser cooling of our beam using a three-frequency serrodyne waveform in further detail. To quantify the cooling, we fit the integrated experimental fluorescence data with a double-Gaussian model and extract the ratio of the amplitudes of the two Gaussians as a measure for the cooling efficiency (see Appendix). We study this ratio as a function of detuning and magnetic field, with all observations in good agreement with our simulations. Notably, in the case of Sisyphus cooling, the detuning that yields the optimal cooling force is several times larger than the natural linewidth of the transition employed. This is in contrast to Doppler forces, which are maximized in a small frequency interval, on the order of the natural linewidth. The larger detunings lead to reduced scattering in Sisyphus cooling. From our data, we estimate a number of $100-200$ photons, compared to a more resonant situation, where many hundreds of photons are scattered. Correspondingly, we find branching losses to be negligible in our implementation of Sisyphus cooling for BaF.

\begin{figure}[tb]
\includegraphics[width=0.47\textwidth]{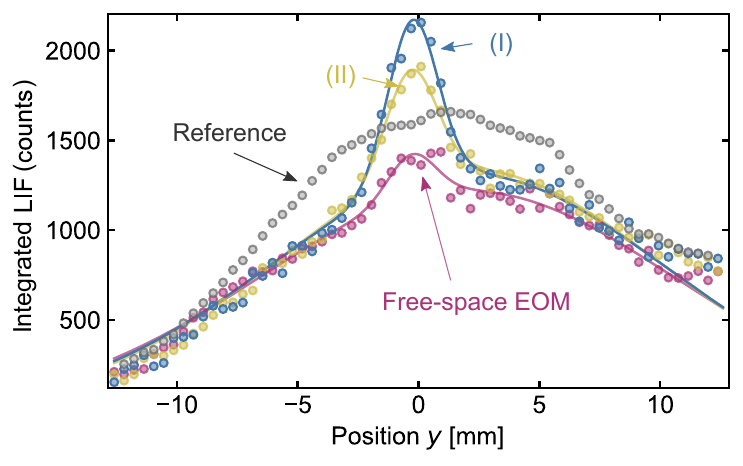}
 \caption{Comparison of the cooling efficiency using a three-frequency serrodyne waveform (scheme I in Fig.~\ref{fig:levelscheme}), four-frequency serrodyne waveform (scheme II in Fig.~\ref{fig:levelscheme}) and a free-space EOM that is sinusoidally driven with a frequency of $39.3\,$MHz to approximately match BaF's ground state hyperfine splitting. For convenience, an additional serrodyne was used in the latter case to precisely shift the laser to the molecular resonance. The reference corresponds to an unperturbed molecular beam, with no laser light applied in the cooling region.}
\label{fig:comparison}
\end{figure}

To further quantify the efficiency of the three-frequency serrodyne approach, in Fig.~\ref{fig:comparison} we compare the achieved cooling with the results of a four-frequency serrodyne waveform (scheme II in Fig.~\ref{fig:levelscheme}), as well as the more conventional, permanent sidebands created using a sinusoidally driven free-space EOM approximately matching the ground state hyperfine splitting. Consistent with our modeling, the three-frequency serrodyne results (scheme I) outperform the others in terms of cooling, while also showing the least amount of losses. 

In conclusion, we have experimentally demonstrated one-dimensional transverse laser cooling of a beam of \baff{138} molecules. Our results have important consequences for future precision measurements with BaF molecules. In Ref.~\cite{Aggarwal2018}, transverse Doppler cooling of a slow BaF beam was proposed to enhance the sensitivity of searches for the electron's electric dipole moment. The observed laser cooling forces in our experiment exceed the ones considered in that work significantly, while requiring far less photon scattering events. Similarly, in Ref.~\cite{Demille2008} odd isotopologues of BaF were proposed as precision sensors for nuclear parity violation. A proof of principle experiment with \baff{138} confirmed the high sensitivity of these molecules for such measurements~\cite{Altuntas2018}. However, a finite parity violation signal is only expected to be present in odd isotopologues like \baff{137} and \baff{135}, which are far less abundant. Realizing similar cooling forces as the ones presented here for these isotopologues, a several order of magnitude increase in molecular numbers could be realized~\cite{Kogel2021}, bringing high-precision studies of nuclear parity violation within reach. Furthermore, as previously observed for YbF molecules~\cite{Alauze2021}, our simulations suggest that extending the cooling to two transverse directions will further increase its phase space acceptance. Moreover, it could be combined with focusing techniques~\cite{Fitch2021methods, Touwen2024} to yield an even larger number of ultracold molecules. 

To our knowledge, our work represents the first use of serrodynes, or variants thereof, in laser cooling beyond radiation pressure forces~\cite{Holland2021}. In the future, it will be an interesting question to use the freedom afforded by the serrodynes to further optimize the cooling parameters, e.g. by identifying the optimal order and strength of individual frequency components given certain molecular level structures. More generally, our approach to generate Sisyphus laser cooling forces using synthesized optical spectra thus opens the door to the laser cooling of molecular species, where a large number of transitions would otherwise render laser cooling using conventional sidebands prohibitively complex. 

\section*{Acknowledgments}
We are indebted to Tilman Pfau for generous support. We thank Ralf Albrecht and Einius Pultinevicius for contributions in the early stage of the experiment, Philipp Flad, Mario Hentschel and Harald Giessen for the production of optical coatings, and Rees McNally and Qi Sun for discussions. This project has received funding from the European Research Council (ERC) under the European Union’s Horizon 2020 research and innovation programme (Grant agreement No. 949431), Vector Stiftung, Carl Zeiss Stiftung, and the RiSC initiative of the Ministry of Science, Research and Arts Baden-W\"urttemberg.

\section*{Appendix}

\begin{figure}[tb]
\includegraphics[width=0.46\textwidth]{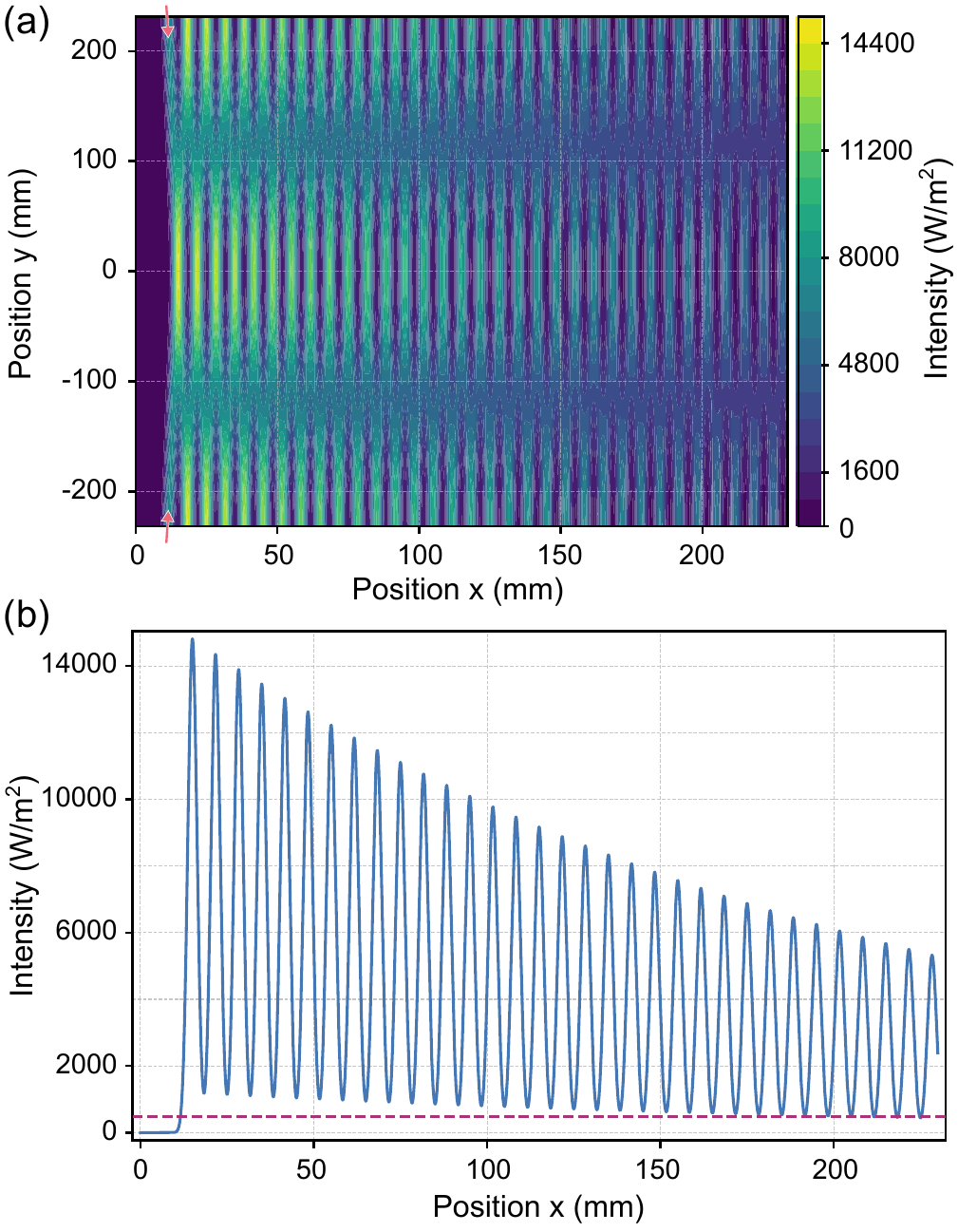}
 \caption{(a) Simulated intensity profile in the cooling region. Each of the two counter-propagating laser beams (indicated by arrows) undergoes $17$ round trips, leading to $34$ intensity maxima where both beams overlap. (b) Cut through the center ($y=0$) of the intensity profile, along the molecular beam propagation axis. The dashed line is $100\times I_{sat}$, which corresponds to a well-saturated molecular multi-level system. Here, $I_{sat} = 5.84\,\mathrm{W/m}^2$ is the saturation intensity of a two-level system with the same transition properties.}
\label{fig:intensity}
\end{figure}

\subsection{Cooling region laser setup}
To obtain the results presented in Fig.~\ref{fig:cooling}, we use light derived from several diode lasers that are amplified by tapered amplifiers to provide ample power for the cooling process. The generation of sidebands for the main cooling laser is implemented using a fiber-coupled EOM, which directly connects the initial diode laser system with its tapered amplifier. In contrast to the main cooling transition, the repumping and depumping transitions only contribute very little to the overall cooling forces and only serve to avoid branching losses or realize imaging, respectively. It is thus sufficient to off-resonantly address the hyperfine substates involved in these transitions using conventional free-space EOMs, tuned to create equidistant sidebands with a spacing of $39.3$\,MHz. 

Custom mirror and vacuum window coatings in the cooling region ensure that the peak intensities of the beams remain at over $30\%$ of the initial value even after $17$ round-trip passes between the mirrors. The full width at half maximum of the beams is $3.1\,$mm. With the powers given in the main text, these parameters correspond to intensities well above saturation for the multi-level molecular system throughout the cooling region. A simulation of the intensity profile is shown in Fig.~\ref{fig:intensity}. 

During cooling, a set of coils allows for the application of a variable, homogeneous magnetic field that forms an angle of $45^\circ$ with the cooling laser polarization direction, which realizes the remixing of molecular dark states. 

We note the very sensitive alignment required to avoid imbalances in the forces, as previously also observed in BaH~\cite{McNally2020} and SrOH molecules~\cite{Kozyryev2017}. Due to BaF's small linewidth, we use transverse Doppler shift measurements to mutually pre-align our beam and the cooling region to around $1^\circ$, and subsequently use pinholes on micrometer stages to improve the beam positioning even further. 

\subsection{Frequency spectra}
Due to the finite bandwidth of our arbitrary waveform generator (AWG) and amplifier, as well as their non-linear response the generation of a serrodyne waveform results not only in a shifted frequency of the laser beam, but also creates a comb of weak higher-order frequency components. If desired, these additional components could be further reduced using a higher bandwidth AWG, as well as a systematic compensation of non-linearities of the whole frequency-shifting setup from the AWG to the amplifier to the EOM~\cite{Holland2021}. 

\begin{figure}[h]
\centering
\includegraphics[width=0.46\textwidth]{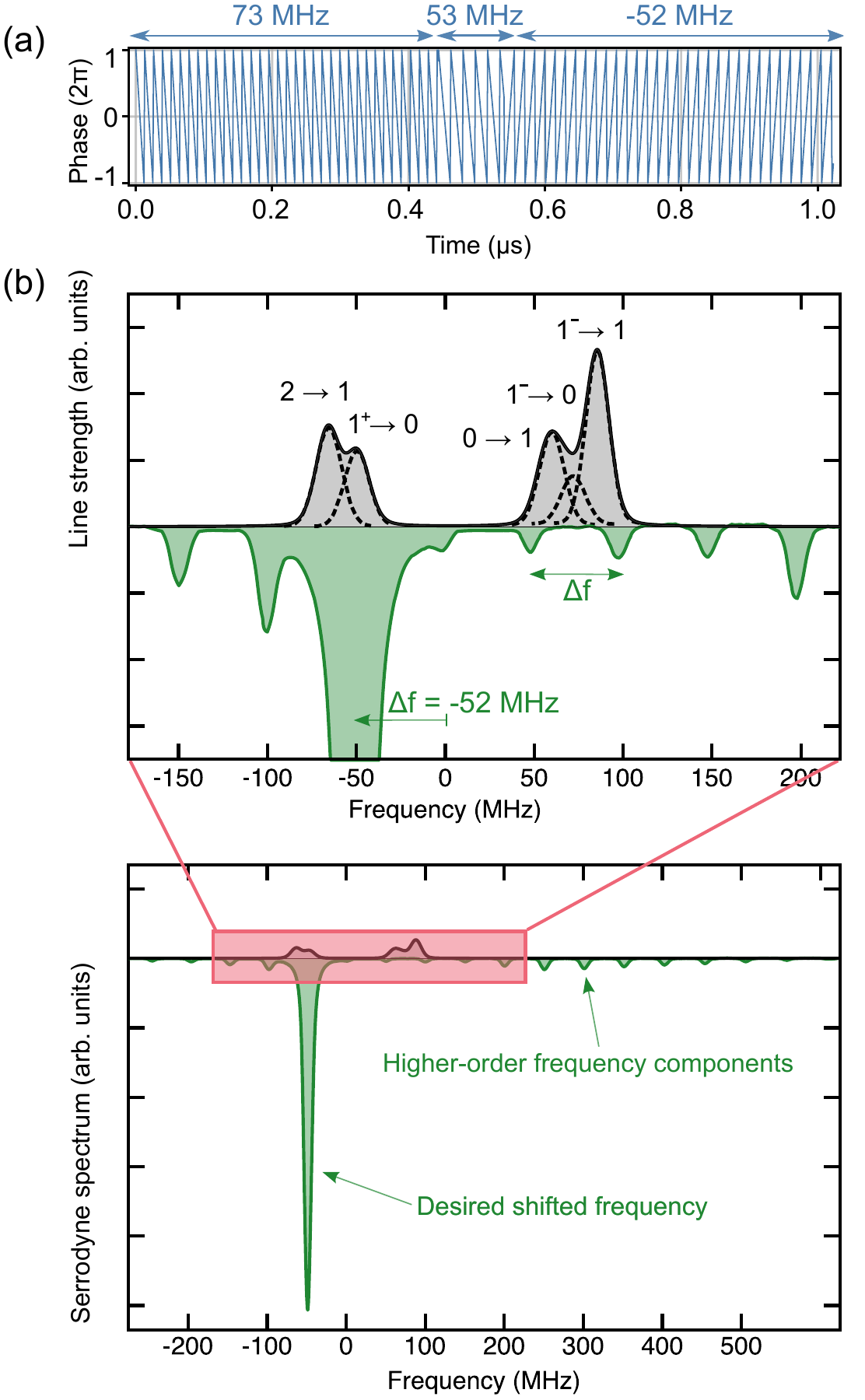}
 \caption{(a) Example of a time-sequenced, three-frequency serrodyne waveform, containing three concatenated linear voltage ramps. (b) Even if only a single linear ramp is applied to shift the frequency of a laser by $\Delta f$, the finite bandwidth of our arbitrary waveform generator, as well as non-linear effects, results in a comb of weak higher-order frequency components, which are equally spaced by $\Delta f$ (bottom, green). The spectrum shown has been recorded using a Fabry-Perot resonator, with a $1.5\,$GHz FSR and a finesse of 200. We intentionally do not compensate these higher-order components as they can off-resonantly drive unaddressed transitions (top, gray) and thus minimize dark states. These components contain around $50\%$ of the total power for our parameters. Similar higher-order frequency components appear for all three linear ramps concatenated in (a). As a consequence, even for time-sequenced serrodynes, dark states are effectively suppressed, leading to strong laser cooling forces.}
\label{fig:serrodynes}
\end{figure}

In this work, instead, we use these higher-order sidebands as a persistent frequency background that is very much independent of the particular serrodyne applied at any given time. It can therefore be used to permanently provide a weak repumping of unaddressed ground-state hyperfine levels, as illustrated in Fig.~\ref{fig:serrodynes}.
In this way, despite the comparably low switching speed between individual serrodyne frequency components, we observe strong Sisyphus forces without significant losses. 

\subsection{Data analysis and peak height ratio}
To characterize the efficiency of the laser cooling we fit our data, both experimental and simulated, using a double-Gaussian model function:
\begin{equation*}
    f(y) = A e^{-\frac{(y-y_{0,A})^2}{2\sigma_A^2}} + B e^{-\frac{(y-y_{0,B})^2}{2\sigma_B^2}}
\end{equation*}
For example fits see Fig.~\ref{fig:cooling}d. From the amplitudes $A$ and $B$ of such fits, we extract the \textit{peak height ratio} $A/B$, which corresponds to the peak height of the coldest molecules, normalized to the uncooled molecular signal. Examples of peak height ratios for different detunings and magnetic fields are given in Fig.~\ref{fig:comparison}. 

\subsection{Laser cooling simulations}
To compare our experimental data to simulations we solve optical Bloch equations to create force profiles as a function of all relevant laser parameters. We then propagate molecules through the experimental apparatus using these force profiles. Our simulations include the $12$ hyperfine levels of the \gs($\nu=0$) ground state, as well as the $4$ hyperfine levels of the \exs($\nu=0$) excited state. Energy splittings and branching ratios are based on the spectroscopic data reported in our previous work~\cite{Rockenhaeuser2023}. Including repumping lasers, and thus more ground states, is known to reduce the magnitude of the laser cooling forces \cite{Kogel2021}. We heuristically take this into account by scaling down all calculated force profiles by $(50\pm10)\,$\%.

As previously observed~\cite{Holland2021}, we find no notable difference between static simulations where all frequency sidebands are permanently present, and time-sequenced serrodynes with high enough switching frequency. All simulation results reported thus represent solutions of the OBEs where we assume all frequencies to be permanently present. A detailed comparison of the various serrodyne configurations possible will be reported in future work.

\bibliography{biblio}

\end{document}